\pdfoutput=1
\documentclass{article}
\usepackage{fullpage}
\usepackage{graphicx}
\usepackage{amsmath, amsfonts, amssymb}
\usepackage{xcolor}
\usepackage{etoolbox}

\newtoggle{icassp}

\title{On The Inductive Bias of Words in Acoustics-to-Word Models}
\author{Hao Tang, James Glass}
\date{Massachusetts Institute of Technology\\
Computer Science and Artificial Intelligence Laboratory\\
Cambridge MA, USA\\
{\small\texttt{\{haotang, glass\}@mit.edu}}}

\begin{document}

\maketitle

\begin{abstract}
Acoustics-to-word models are end-to-end speech recognizers
that use words as targets without relying on pronunciation
dictionaries or graphemes.  These models are
notoriously difficult to train due to the lack of linguistic
knowledge.  It is also unclear how the amount
of training data impacts the optimization and generalization
of such models.  In this work, we study
the optimization and generalization of acoustics-to-word
models under different amounts of training data.
In addition, we study three types of inductive bias,
leveraging a pronunciation dictionary, word boundary
annotations, and constraints on word durations.
We find that constraining word durations leads
to the most improvement.   Finally, we analyze the
word embedding space learned by the model,
and find that the space has a structure
dominated by the pronunciation of words.
This suggests that the contexts of words,
instead of their phonetic structure, should be
the future focus of inductive bias in acoustics-to-word models.
\end{abstract}

\section{Introduction}

Acoustics-to-word models are a special class of end-to-end models
for automatic speech recognition (ASR) where the output targets are words~\cite{SLS2017, A+2017, L+2017,
L+2018, A+2018, UIMK2018, Y+2018, PM2018}.
In contrast to other end-to-end models where the output
targets are phonemes or graphemes~\cite{MGM2015, CJLV2016}, acoustics-to-word models
directly predict words without relying on any intermediate lexical units.
A parallel set of transcriptions (in words) and acoustic recordings
is sufficient to train these models.
This property offers a significant edge over conventional
ASR systems, because
acoustics-to-word models require minimal domain
expertise to train and use, and might potentially be cheaper to build than conventional
speech recognizers, depending on the cost of experts and obtaining
the necessary resources.

When not given enough data, acoustics-to-word models
are notoriously difficult to train~\cite{A+2017}.
The reason behind the difficulty is often attributed
to the lack of inductive bias, e.g., linguistic knowledge about phonemes
and the pronunciation of words.
To improve the performance of acoustics-to-word models,
much of the previous work has focused on injecting inductive bias
to the models, such as initializing acoustics-to-word models with
a pre-trained phone recognizer \cite{A+2017, A+2018, Y+2018}.
This approach has been shown to be critical in training
acoustics-to-word models.
However, it defeats the purpose of acoustics-to-word models
by requiring a lexicon during training.

The central question is whether a lexicon is really necessary
in training acoustics-to-word models.
The result in~\cite{A+2017} suggests that
the optimization of acoustics-to-word models is inherently difficult,
and they further argue that fitting the training set itself is difficult
because there is not enough data.
Additional inductive bias can provide a better initialization
for optimizing the objective.
However, it is unclear and rather counter-intuitive
that fitting a small training set is more difficult
than fitting a large one.

In this work, we first study the optimization of acoustics-to-word
models.  We show that, in contrast to previous studies,
acoustics-to-word models are able fit a training set of various sizes
without any inductive bias.  We then study the sample complexity of
acoustics-to-word models, and show that, instead of having
an all-or-none learning effect as seen in~\cite{A+2017},
the generalization error decreases as we increase the number of samples.
In addition, we study the acoustics-to-word models under
different inductive biases, including the pronunciation of words,
word boundaries, and word durations.
We also vary the amount of additional data needed for
these inductive biases.
These results characterize the optimization
and generalization of acoustics-to-word models.

Given the ability to train an acoustics-to-word model end to end,
it is natural to ask to what extent the common resources, such as the
phoneme inventory, pronunciation dictionary, and language model,
are learned by the model.
We analyze the weights in the last layer for predicting words
and find special structures in the word embedding space.
These results show the limitation of acoustics-to-word models
and shed light on future direction for improvement.

\section{Acoustics-to-Word Models}

The class of acoustics-to-word models can be either implemented
with connectionist temporal classification (CTC) \cite{GFGS2006}
or sequence-to-sequence models \cite{PM2018}.
In this work, we focus on the case of CTC acoustics-to-word models.

Let $x = (x_1, \dots, x_T)$ be an input sequence of $T$ acoustic feature
vectors, or frames, where $x_t \in \mathbb{R}^d$ for
$t = 1, \dots, T$.  Let $y = (y_1, \dots, y_K)$ be an output sequence
of $K$ labels, where $y_k \in L$ for $k = 1, \dots, K$ and a vocabulary set $L$.
For example, in our case, $x_t$ is a log-Mel feature vector,
and $y_k$ is word.
We use the $|\cdot|$ to denote the length of a sequence.
For example, in this case, $|x| = T$ and $|y| = K$.
The goal of the model is to map an input sequence to
an output sequence.

In ASR, it is typical to have an output sequence
shorter than the input sequence, i.e., $K < T$.
In order to predict $K$ labels given $T$ frames,
a CTC model predicts $T$ labels, allowing the prediction
to have repeating labels and $\varnothing$'s,
where $\varnothing$ is the blank symbol for predicting nothing.
To get the final output sequence, there is a post-processing
step $B$ that removes the repeating labels and the blank
symbols, in that order.
To train a CTC model, we maximize the conditional likelihood
\begin{align}
p(y | x) = \sum_{z \in B^{-1}(y)} \prod_{t=1}^T p(z_t | x_t)
\end{align}
for a sample pair $(x, y)$, where
$B^{-1}$ maps a label sequence of length $|y|$ to the set of all possible
sequences of length $|x|$ by repeating labels and inserting $\varnothing$'s.
In other words, the function $B^{-1}$ the pre-image of $B$,
and note that $z_t \in L \cup \{\varnothing\}$ for $t = 1, \dots, |x|$.
The conditional likelihood
and its gradient with respect to each individual $p(z_t | x_t)$
can be efficiently computed with dynamic programming \cite{GFGS2006}.

The individual conditional probability $p(z_t | x_t)$ is typically
modeled by a neural network.  The input sequence $x_1, \dots, x_T$
is first transformed into a sequence of hidden vectors
$h_1, \dots, h_T$, and $p(z_t | x_t) = \log \text{softmax}(W h_t)$
for $t = 1, \dots, T$.  Different studies have used
different network architectures \cite{MGM2015}.
In this work, we use long short-term
memory networks (LSTMs) to transform the input sequence.

\section{Inductive Bias of Words}

Given how little domain knowledge is used in acoustics-to-word models,
much effort has been put into injecting inductive bias
into acoustics-to-word models.
In this section, we describe three approaches to achieve this.

\subsection{Lexicon}

A lexicon provides a mapping from a word
to its canonical pronunciations.  In conventional speech recognizers,
hidden Markov models are constructed for each phoneme
(under different contexts).
Once the models are trained,
it is straightforward to generalize to unseen words by
adding the pronunciations of those words to the lexicon.

In end-to-end speech recognizers, the need for a lexicon
is sidestepped by using graphemes instead phonemes
as targets.  Some end-to-end speech recognizers are able
to generalize to unseen words this way, but it is still very
difficult to add or remove a word from the vocabulary of
an end-to-end speech recognizer.  In terms of acoustics-to-word
models, it is easy to remove a word from the vocabulary,
but hard for the models to generalize to unseen words
\cite{L+2018}.

To make use of a lexicon, past studies
train a phoneme-based CTC model on the phoneme sequences
converted from word sequences in the training set \cite{A+2017, A+2018, Y+2018}.
Acoustics-to-word models are
then initialized with the pre-trained phoneme-based CTC model
with the hope that the acoustics-to-word models
are able to utilize the phonetic knowledge encoded in
the phoneme-based CTC model.

\iftoggle{icassp}{\vspace{-0.3cm}}{}
\subsection{Word Boundary}

Another type of inductive bias is word boundaries.
Speech recognizers in general tend to perform worse on long utterances
than short ones, and rely on insertion penalties
to correct such bias.
The training errors of long utterances are also typically
higher than those of short ones.
One hypothesis is that word boundaries are more difficult
to pinpoint during training for long utterances than for short ones.
In similar spirit, if we have access to word boundaries,
we can train a frame classifier to encode the knowledge,
and transfer it to acoustics-to-word models through
initialization.


\iftoggle{icassp}{\vspace{-0.3cm}}{}
\subsection{Word Duration}

Another factor that makes pinpointing word boundaries difficult
is the high variance of word durations.
In the WSJ training set (\texttt{si284}),
the average duration of a word (measured from the forced alignments)
is 339.5 ms with a standard
deviation of 198.0 ms, while the average duration of a phoneme
is 81.6 ms with a standard deviation of 46.7 ms.
These statistics show that it is more difficult to
estimate the number of words in an utterances than
the number of phonemes.

To introduce word duration bias in acoustics-to-word models,
we down-sample the hidden vectors after each LSTM layer.
Specifically, suppose $h^{n}_1, \dots, h^{n}_T$ is the
output vectors produced by the $(n-1)$-th LSTM layer after taking
$h^{n-1}_1, \dots, h^{n-1}_T$ as input.
Instead of the entire $T$ vectors, we feed
$h^{n}_1, h^{n}_3, \dots, h^{n}_{2 \lfloor T/2 \rfloor -1}$
to the $n$-th LSTM layer.
Every down-sampling reduces the frame rate by half.
Down-sampling has been introduced in the past for speeding
up inference \cite{VMH2013, CJLV2016, M+2016, TWKL2016},
but it can also act as imposing a constraint on
the minimum word duration \cite{T+2017}.

\section{Experiments}

We choose the Wall Stree Journal data set (\texttt{WSJ0} and \texttt{WSJ1})
for our experiments for the wide variety of words
and rich word usage in the data set.  It consists of 80 hours of
read speech and 13,635 unique words in the training
set (\texttt{si284}).  We follow the standard protocol using 90\% of the
\texttt{si284} for training, 10\% of \texttt{si284} for development,
and \texttt{dev93} and \texttt{eval92} for testing.
We obtain forced alignments of words and phonemes with
a speaker-adaptive Kaldi system, following the standard recipe \cite{P+2011}.
We extract 80-dimensional log-Mel features,
and use them as input without concatenating i-vectors.
We use a 4-layer unidirectional LSTM with 500 units per layer.
The softmax targets at each time step are the 13,635 words
plus the blank symbol ($\varnothing$).

The LSTMs are trained by minimizing the CTC loss with
vanilla stochastic gradient descent (SGD)
for 20 epochs, a step size of 0.05,
and gradient clipping of norm 5.  The mini-batch
size is one utterance.
The best model within the 20 epochs is selected and trained
for another 20 epochs with learning rate 0.0375
decayed by 0.75 after each epoch.
The best model out of the 40 epochs is used for evaluation.
No additional regularization is used.

To see how the amount of data impacts optimization and generalization,
we train the LSTMs on different amount of training data,
one-half of \texttt{si84} (around 5 hours),
\texttt{si84} (around 10 hours),
one-half of \texttt{si284} (around 35 hours)
and the entire \texttt{si284} (around 70 hours).
Results are shown in Table~\ref{tbl:baseline}.
We measure the training perplexity of the last epoch, i.e.,
the sum of cross entropy at each frame divided by the number
labels (as opposed to the number of frames).
Except for the one with half of \texttt{si84},
the other three models are able to fit the training set
without trouble.
Note that the models receive different numbers of
gradient updates, so it is not surprising
that lower training error is observed when using more data.
The generalization error is
better when using more data as expected.

To introduce the inductive bias given by the lexicon,
we train a 3-layer LSTM phoneme-based CTC model using the same
training procedure.
The quality of the phoneme recognizer is shown in Table~\ref{tbl:phone-ctc},
and the performance is on par with the state of the art 
of a similar architecture \cite{MGM2015}.
Similarly, to introduce the inductive bias of word boundaries,
we train a 4-layer LSTM word frame classifier
with a lookahead of one frame and online decoding~\cite{TG2018}.
The quality of the word frame classifier is shown in Table~\ref{tbl:word-frame}.
This architecture with additional lookahead is able to achieve
state-of-the-art frame error rates~\cite{TG2018}.
However, to limit the confounding factors, we fix the lookahead to one.

To see the impact of the amount of training data,
we initialize the acoustics-to-word models with
phoneme-based CTC models and word frame classifiers trained
on different amounts of training data.
The bottom three layers of LSTMs are initialized
with the various pre-trained models.
In other words, the last layer of the frame
classifier is discarded.
The last layer and the softmax layer of the initialized acoustics-to-word model
is randomly initialized the same way as the baseline models were.
This number of layers for initialization is motivated by
recent success in multitask CTC \cite{TTLK2017, SM2018}.
As a comparison, we also have a set of acoustics-to-word models
initialized with the bottom three layers of
the trained acoustics-to-word models in Table~\ref{tbl:baseline}.
Results are shown in Table~\ref{tbl:warm}.
Not only do we see no improvement in initializing with phoneme-based CTC models
and frame classifiers, it actually hurts performance in many cases.
Small improvements from initializing with acoustics-to-word models
themselves has been observed, with the best model achieving 27.5\%
WER on \texttt{dev93} and 24.4\% on \texttt{eval92}.

Finally, we experiment with the amount of down-sampling in LSTM layers.
We have four LSTMs with increasing amount of down-sampling
after the input.  Results are shown in Table~\ref{tbl:downsampling}.
We see significant improvement, with
the best having a down-sampling factor of four.
This is consistent with the commonly used frame rate \cite{VMH2013, M+2016, TWKL2016, T+2017}.

\begin{table}
\caption{\label{tbl:baseline} WERs (\%) of acoustics-to-word models
    trained on different sizes of the training set.
    The last column is the training perplexity of the last epoch.}
\iftoggle{icassp}{\vspace{-0.1cm}}{}
\begin{center}
\begin{tabular}{lrrr||r}
train               & dev   & \texttt{dev93} & \texttt{eval92} & \texttt{si284} \\
\hline
\texttt{si84}-half  & 63.0  & 64.8           & 58.3            & 1.27 \\
\texttt{si84}       & 51.9  & 55.1           & 45.1            & 0.54 \\
\texttt{si284}-half & 32.9  & 36.2           & 33.5            & 0.40 \\
\texttt{si284}      & 21.4  & 29.4           & 26.3            & 0.34
\end{tabular}
\end{center}
\iftoggle{icassp}{\vspace{-0.3cm}}{}
\end{table}

\begin{table}
\caption{\label{tbl:phone-ctc} PERs (\%) of phoneme-based CTC models
    trained on different sizes of the training set.}
\iftoggle{icassp}{\vspace{-0.1cm}}{}
\begin{center}
\begin{tabular}{lrrr}
train               & dev   & \texttt{dev93} & \texttt{eval92} \\
\hline
\texttt{si84}-half  & 38.0  & 40.0           & 26.1            \\
\texttt{si84}       & 27.7  & 29.1           & 16.1            \\
\texttt{si284}-half & 16.1  & 15.0           & 12.4            \\
\texttt{si284}      & 12.3  & 11.9           &  9.4
\end{tabular}
\end{center}
\iftoggle{icassp}{\vspace{-0.7cm}}{}
\end{table}

\begin{table}
\caption{\label{tbl:word-frame} FERs (\%) of word frame classifiers
    trained on different sizes of the training set.}
\iftoggle{icassp}{\vspace{-0.1cm}}{}
\begin{center}
\begin{tabular}{lrrr}
train               & dev   & \texttt{dev93} & \texttt{eval92} \\
\hline
\texttt{si84}-half  & 67.4  & 62.4           & 61.7            \\
\texttt{si84}       & 56.0  & 55.3           & 53.2            \\
\texttt{si284}-half & 40.6  & 46.3           & 47.2            \\
\texttt{si284}      & 28.3  & 44.3           & 45.7
\end{tabular}
\end{center}
\end{table}

\begin{table}
\caption{\label{tbl:warm} WERs (\%) on the development set for acoustics-to-word models
    initialized from different models.  The first column indicates the
    amount of data on which the initial models are trained.}
\iftoggle{icassp}{\vspace{-0.1cm}}{}
\begin{center}
\begin{tabular}{lrrr}
                    & phoneme CTC & word frame   & word CTC \\
\hline                                              
\texttt{si84}-half  & 27.4        & 25.9         & 22.3  \\
\texttt{si84}       & 28.9        & 24.8         & 21.9  \\
\texttt{si284}-half & 25.0        & 31.1         & 20.6  \\
\texttt{si284}      & 21.1        & 22.0         & 19.4 
\end{tabular}
\end{center}
\iftoggle{icassp}{\vspace{-0.8cm}}{}
\end{table}

\begin{table}
\caption{\label{tbl:downsampling} WERs (\%) of acoustics-to-word models for different
     amount of down-sampling in the LSTM layers.}
\iftoggle{icassp}{\vspace{-0.1cm}}{}
\begin{center}
\begin{tabular}{lrrr}
down-sampling  & dev   & \texttt{dev93} & \texttt{eval92} \\
\hline
$2^0$         & 21.4  & 29.4           & 26.3            \\
$2^1$         & 21.8  &                &                 \\
$2^2$         & 17.7  & 26.8           & 24.6            \\
$2^3$         & 18.6  &                &                 \\
$2^4$         & 18.8  &                &
\end{tabular}
\end{center}
\iftoggle{icassp}{\vspace{-0.8cm}}{}
\end{table}

\section{Analysis}

The gap between the results in Table~\ref{tbl:baseline} and the
state-of-the-art end-to-end system~\cite{ZCJ2017},
i.e., 14.1\% absolute on \texttt{eval92}, is large,
and it shows that a vanilla acoustics-to-word model
does not perform well out of the box.
A phoneme-based CTC model with a lexicon
and without a language model can achieve 26.9\% WER on
\texttt{eval92} \cite{MGM2015}.
Based on these results and the best down-sampling
factor, it is likely that the model only learns to map acoustic features
to words without utilizing the dependencies between words much.

\begin{figure}[t]
\begin{center}
\includegraphics[width=5.5cm]{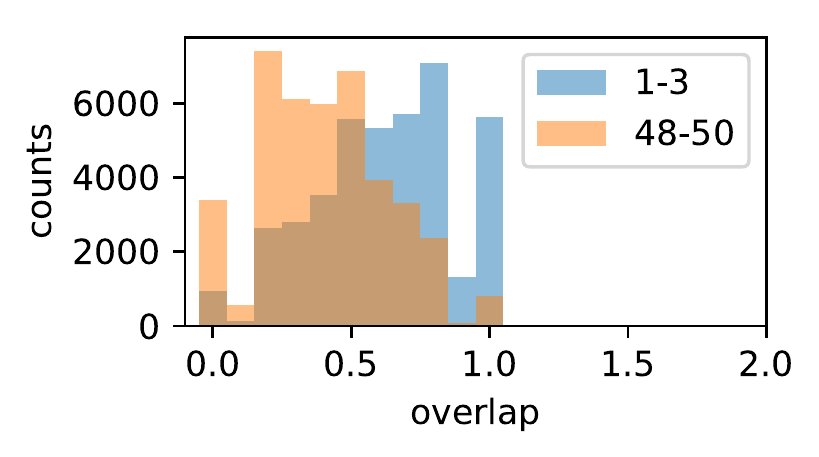}
\iftoggle{icassp}{\vspace{-0.4cm}}{}
\caption{\label{fig:pron-edit}
    The histogram of the amount of overlap (normalized by the number of phonemes in the shorter word)
    in canonical pronunciations of a word
    to its close neighbors (the first to the third nearest neighbor) and far neighbors
    (the 48th to the 50th nearest neighbor).}
\end{center}
\iftoggle{icassp}{\vspace{-0.5cm}}{}
\end{figure}

To confirm this hypothesis and to understand how words
are arranged in the embedding space, we analyze the weights of the
softmax layer.
Given a word, we take its corresponding weight vector in the softmax layer
and compute its nearest neighbors.  We first notice that
similar pronouncing words tend to be nearest neighbors of each other.
To see this, we define the overlap (in canonical pronunciation)
as the number of phoneme tokens appeared in both words divided by
the length of the shorter word.  We compare the overlaps of words
from two groups, the word to its close neighbors (the first
to the third neighbor) and the word
to its far neighbors (the 48th to the 50th neighbor).
As shown in Fig.~\ref{fig:pron-edit}, the close neighbors have
more similar pronunciations than the far neighbors.
This confirms the hypothesis that the model relies more
on the acoustics rather than the language to predict words.

\begin{figure}[t]
\begin{center}
\includegraphics[width=6cm]{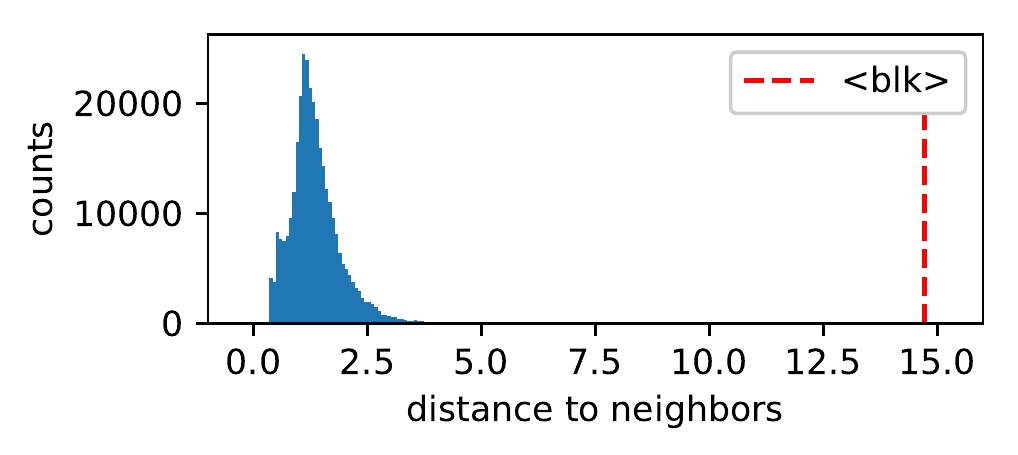}
\iftoggle{icassp}{\vspace{-0.4cm}}{}
\caption{\label{fig:blk} The histogram of distances from a word to its top 25 nearest neighbors.
    The average distance of \texttt{<blk>} (i.e., $\varnothing$) to its
    nearest neighbors is shown in dashed red.}
\end{center}
\iftoggle{icassp}{\vspace{-0.5cm}}{}
\end{figure}

\begin{figure}[tb]
\iftoggle{icassp}{\vspace{-0.2cm}}{}
\begin{center}
\includegraphics[width=6cm]{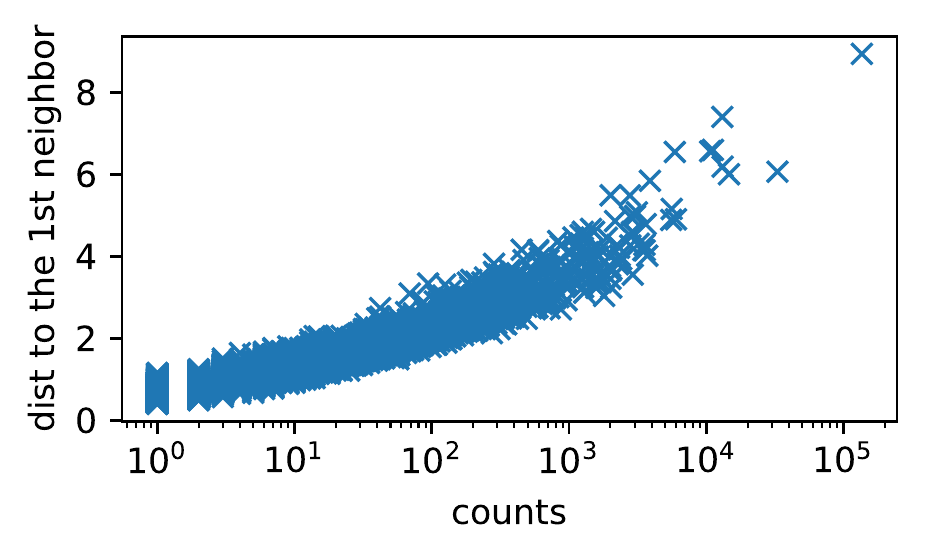}
\iftoggle{icassp}{\vspace{-0.4cm}}{}
\caption{\label{fig:margin} The number of times a word appears in the training set against
    the distance to its first nearest neighbor.  The point in the upper
    right corner is \texttt{SIL} (the silence).}
\end{center}
\end{figure}

We then notice that the blank symbol is significantly further
away (in L2 distance) from other words, and this can be seen from the histogram
of distances in Fig.~\ref{fig:blk}.
We refer to the distance between a word and its first nearest neighbor
as the margin.  In other words, the margin of the blank symbol is
significantly larger than that of other words.
We then notice that the margin is related to the occurrence of words
in the training set, as is shown in Fig.~\ref{fig:margin}.
This might be the consequence of conditional independence
assumed in the CTC loss. If this is indeed the case,
improved performance can only be achieved through
a different objective or an inductive bias on word dependency.

\section{Conclusion}

In this work, we study the optimization and generalization
of acoustics-to-word models.  We find that the models
are able to fit data sets of various sizes
without trouble.  The generalization error decreases as
expected, as we increase the amount of training data.
In contrast to other studies, we find no improvement
in initializing the models with
a pre-trained phoneme-based CTC model or a word frame classifier.
Down-sampling the hidden vectors after the LSTM layer
provides significant improvement.
To understand what hinders the performance,
we analyze the word embeddings learned by the model.
The model discovers
similar sounding words and places them in the corresponding
neighborhood.  However, the model might not be utilizing
the label dependency much during decoding.
This suggests that label dependency should be the future
focus of inductive bias in acoustics-to-word models.

\bibliographystyle{plain}
\bibliography{wordbias}

\end{document}